\documentclass[12pt]{article}
\usepackage{graphicx}
\usepackage{ifthen} 
\newboolean{pdflatex}
\setboolean{pdflatex}{true} 

\newboolean{articletitles}
\setboolean{articletitles}{true} 

\newboolean{uprightparticles}
\setboolean{uprightparticles}{false} 

\usepackage{microtype}
\usepackage{lineno}  
\usepackage{xspace} 

\usepackage{graphicx}  
\usepackage{color}
\usepackage{colortbl}

\usepackage{amsmath} 
\usepackage{amssymb}
\usepackage{amsfonts}
\usepackage{bm}
\usepackage{upgreek} 

\def\pbnr{}
\def\speaker{Liang Sun}
\def\onbehalfof{\lhcb Collaboration}
\def\title{Two-body wrong-sign mixing and \CP violation}
\def\affiliation{Department of Physics\\
University of Cincinnati, Cincinnati, Ohio 45221-0011 USA}

\newcommand\pubnumber{\pbnr}
\newcommand\pubdate{\today}
\def\Title#1{\begin{center} {\Large #1 } \end{center}}
\def\Author#1{\begin{center}{ \sc #1} \end{center}}

\newcommand{\OnBehalf}[1]{\sbox0{#1}\ifdim\wd0=0pt
        {}
	\else
	{\\on behalf of #1}
	\fi}
\newcommand{\SupportedBy}[1]{\sbox0{#1}\ifdim\wd0=0pt
        {}
	\else
	{\footnote{#1}}
	\fi}
\def\Address#1{\begin{center}{ \it #1} \end{center}}

\newcommand\pubblock{\includegraphics[width=5cm]{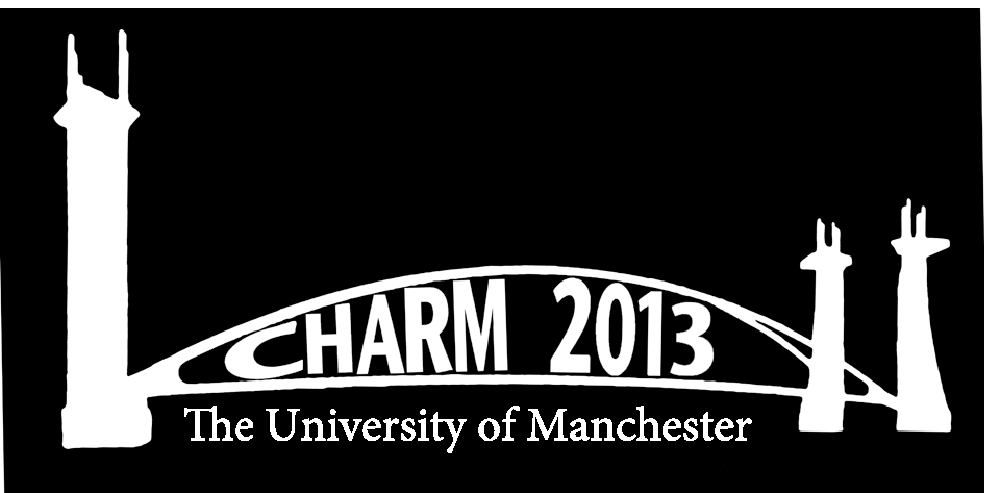}\hfill{\begin{tabular}{l} \pubnumber\\
         \pubdate  \end{tabular}}}
\newenvironment{Abstract}{\begin{quotation}  }{\end{quotation}}
\newenvironment{Presented}{\begin{quotation} \begin{center} 
             PRESENTED AT\end{center}\bigskip 
      \begin{center}\begin{large}}{\end{large}\end{center} \end{quotation}}
\def\Acknowledgements{\bigskip  \bigskip \begin{center} \begin{large}
             \bf ACKNOWLEDGEMENTS \end{large}\end{center}}
\def\venue{The 6$^{th}$ International Workshop on Charm Physics\\
(CHARM 2013)\\
Manchester, UK,  31 August -- 4 September, 2013}

\def\beq{\begin{equation}}
\def\eeq#1{\label{#1}\end{equation}}
\def\eeqn{\end{equation}}

\def\beqa{\begin{eqnarray}}
\def\eeqa#1{\label{#1}\end{eqnarray}}
\def\eeqan{\end{eqnarray}}

\let\bar=\overbar

\def\D{{\cal D}}

\def\Dslash{\not{\hbox{\kern-4pt $D$}}}
\def\dslash{\not{\hbox{\kern-2pt $\del$}}}

\def\msb{{\bar{\ssstyle M \kern -1pt S}}}

\def\lhcb {\mbox{LHCb}\xspace}
\def\ux85 {\mbox{UX85}\xspace}

\ifthenelse{\boolean{uprightparticles}}%
{

 \def\Ppi         {\ensuremath{\uppi}\xspace}

 \def\PDelta      {\ensuremath{\Delta}\xspace}                 
 \def\PXi      {\ensuremath{\Xi}\xspace}                 
 \def\PLambda      {\ensuremath{\Lambda}\xspace}                 
 \def\PSigma      {\ensuremath{\Sigma}\xspace}                 
 \def\POmega      {\ensuremath{\Omega}\xspace}                 
 \def\PUpsilon      {\ensuremath{\Upsilon}\xspace}                 
 
 \def\PB      {\ensuremath{\mathrm{B}}\xspace}                 
                  
 \def\PD      {\ensuremath{\mathrm{D}}\xspace}

 \def\PK      {\ensuremath{\mathrm{K}}\xspace}

 \def\Pi      {\ensuremath{\mathrm{i}}\xspace}

}
{

 \def\Ppi         {\ensuremath{\pi}\xspace}

 \mathchardef\PDelta="7101
 \mathchardef\PXi="7104
 \mathchardef\PLambda="7103
 \mathchardef\PSigma="7106
 \mathchardef\POmega="710A
 \mathchardef\PUpsilon="7107
                  
 \def\PB      {\ensuremath{B}\xspace}                 
                  
 \def\PD      {\ensuremath{D}\xspace}

 \def\PK      {\ensuremath{K}\xspace}

 \def\Pi      {\ensuremath{i}\xspace}

}

\def\pion  {\ensuremath{\Ppi}\xspace}

\def\pip   {\ensuremath{\pion^+}\xspace}
\def\pim   {\ensuremath{\pion^-}\xspace}

\def\pipm  {\ensuremath{\pion^\pm}\xspace}
\def\pimp  {\ensuremath{\pion^\mp}\xspace}

\def\kaon  {\ensuremath{\PK}\xspace}
  \def\Kbar  {\kern 0.2em\overline{\kern -0.2em \PK}{}\xspace}

\def\Kz    {\ensuremath{\kaon^0}\xspace}
\def\Kzb   {\ensuremath{\Kbar^0}\xspace}
\def\KzKzb {\ensuremath{\Kz \kern -0.16em \Kzb}\xspace}
\def\Kp    {\ensuremath{\kaon^+}\xspace}
\def\Km    {\ensuremath{\kaon^-}\xspace}
\def\Kpm   {\ensuremath{\kaon^\pm}\xspace}
\def\Kmp   {\ensuremath{\kaon^\mp}\xspace}
\def\KpKm  {\ensuremath{\Kp \kern -0.16em \Km}\xspace}
\def\KS    {\ensuremath{\kaon^0_{\rm\scriptscriptstyle S}}\xspace}

  \def\Dbar    {\kern 0.2em\overline{\kern -0.2em \PD}{}\xspace}
\def\D       {\ensuremath{\PD}\xspace}

\def\Dz      {\ensuremath{\D^0}\xspace}
\def\Dzb     {\ensuremath{\Dbar^0}\xspace}
\def\DzDzb   {\ensuremath{\Dz {\kern -0.16em \Dzb}}\xspace}
\def\Dp      {\ensuremath{\D^+}\xspace}
\def\Dm      {\ensuremath{\D^-}\xspace}
\def\Dpm     {\ensuremath{\D^\pm}\xspace}

\def\DpDm    {\ensuremath{\Dp {\kern -0.16em \Dm}}\xspace}
\def\Dstar   {\ensuremath{\D^*}\xspace}

\def\Dstarp  {\ensuremath{\D^{*+}}\xspace}
\def\Dstarm  {\ensuremath{\D^{*-}}\xspace}

  \def\Bbar    {\kern 0.18em\overline{\kern -0.18em \PB}{}\xspace}

  \def\Y#1S{\ensuremath{\PUpsilon{(#1S)}}\xspace}

\def\Lbar {\ensuremath{\kern 0.1em\overline{\kern -0.1em\PLambda}}\xspace}

\def\to                 {\ensuremath{\rightarrow}\xspace}

\def\CP                {\ensuremath{C\!P}\xspace}

\def\AT#1     {\ensuremath{A_{\mathrm{T}}^{#1}}\xspace}           

\def\C#1      {\ensuremath{\mathcal{C}_{#1}}\xspace}                       
\def\Cp#1     {\ensuremath{\mathcal{C}_{#1}^{'}}\xspace}                    
\def\Ceff#1   {\ensuremath{\mathcal{C}_{#1}^{\mathrm{(eff)}}}\xspace}        
\def\Cpeff#1  {\ensuremath{\mathcal{C}_{#1}^{'\mathrm{(eff)}}}\xspace}       
\def\Ope#1    {\ensuremath{\mathcal{O}_{#1}}\xspace}                       
\def\Opep#1   {\ensuremath{\mathcal{O}_{#1}^{'}}\xspace}                    

\newcommand{\tev}{\ensuremath{\mathrm{\,Te\kern -0.1em V}}\xspace}
\newcommand{\gev}{\ensuremath{\mathrm{\,Ge\kern -0.1em V}}\xspace}
\newcommand{\mev}{\ensuremath{\mathrm{\,Me\kern -0.1em V}}\xspace}
\newcommand{\kev}{\ensuremath{\mathrm{\,ke\kern -0.1em V}}\xspace}
\newcommand{\ev}{\ensuremath{\mathrm{\,e\kern -0.1em V}}\xspace}
\newcommand{\gevc}{\ensuremath{{\mathrm{\,Ge\kern -0.1em V\!/}c}}\xspace}
\newcommand{\mevc}{\ensuremath{{\mathrm{\,Me\kern -0.1em V\!/}c}}\xspace}
\newcommand{\gevcc}{\ensuremath{{\mathrm{\,Ge\kern -0.1em V\!/}c^2}}\xspace}
\newcommand{\gevgevcccc}{\ensuremath{{\mathrm{\,Ge\kern -0.1em V^2\!/}c^4}}\xspace}
\newcommand{\mevcc}{\ensuremath{{\mathrm{\,Me\kern -0.1em V\!/}c^2}}\xspace}

\def\invfb   {\ensuremath{\mbox{\,fb}^{-1}}\xspace}

\def\gsim{{~\raise.15em\hbox{$>$}\kern-.85em
          \lower.35em\hbox{$\sim$}~}\xspace}
\def\lsim{{~\raise.15em\hbox{$<$}\kern-.85em
          \lower.35em\hbox{$\sim$}~}\xspace}

\def\tell1  {TELL1\xspace}
\def\ukl1   {UKL1\xspace}

\newcommand{\massgev}{\gevcc}
\newcommand{\massmev}{\mevcc}

\newcommand{\pis}{\ensuremath{\pi_{\rm s}}\xspace}
\newcommand{\MM}{\ensuremath{M(\Dz\pis^+)}\xspace}

\begin{document}
\begin{titlepage}
\pubblock

\vfill
\Title{\title}
\vfill
\Author{\speaker\OnBehalf{\onbehalfof}}
\Address{\affiliation}
\vfill
\begin{Abstract}
We describe \lhcb measurements for $\Dz$ --\Dzb mixing parameters and 
searches for \CP violation using ``wrong-sign''
$\Dz\to K^+\pi^-$ two-body decays. 
    \lhcb provides the world's most precise 
    measurements of the mixing parameters to date, using
    3~\invfb of $pp$ collision data. By measuring the
    mixing parameters separately for \Dz and \Dzb~mesons, and allowing for \CP violation,
    the \lhcb results also place the world's most stringent constraints 
    on the \CP violation parameters, $|q/p|$ and $A_D$, from a single experiment. 
\end{Abstract}
\vfill
\begin{Presented}
\venue
\end{Presented}
\vfill
\end{titlepage}
\def\thefootnote{\fnsymbol{footnote}}
\setcounter{footnote}{0}
%

\section{Introduction}
For neutral charm mesons, their mass eigenstates are not the same as the their flavor eigenstates, 
    and the difference in mass and width of the two mass eigenstates results in \Dz --\Dzb mixing or oscillation. 
Conventionally, the \Dz mass eigenstates are related to their flavor eigenstates in the linear forms: 
$|D_{1,2} \rangle \equiv p | \Dz \rangle \pm  q|\Dzb \rangle$, where $p$ and $q$ are complex parameters. 
We have the dimensionless mixing parameters
based on the mass and width differences: $x \equiv (m_2 - m_1)/\Gamma$, 
$y \equiv (\Gamma_2-\Gamma_1)/2\Gamma$, where $\Gamma \equiv (\Gamma_2+\Gamma_1)/2$ is the average width. 
In the standard model (SM), very small \Dz mixing is expected with $x, y$ at the 1\% level or less~\cite{Artuso:2008vf}. 
Allowing for \CP violation, which is expected to be very small
in the charm sector, the oscillation rates for mesons produced as \Dz and \Dzb can also differ.
\Dz --\Dzb oscillation occurs through long-distance or short-distance weak processes~\cite{Artuso:2008vf,Bianco:2003vb,Burdman:2003rs}.
Short-distance processes involve flavor-changing neutral currents, and are highly
suppressed in the SM. 
 However physics beyond the SM might come into play 
and alter the average oscillation rate or the difference between \Dz and \Dzb meson rates. 
Studying \CP violation in \Dz oscillation provides an important probe for possible dynamics beyond the SM~\cite{Blaylock:1995ay,Petrov:2006nc,Golowich:2007ka,Ciuchini:2007cw}.


\section{\Dz--\Dzb mixing with wrong-sign $\Dz \to K^+ \pi^-$  decays}

Experimentally, we study right-sign (RS) $\Dz \to K^-\pi^+$ and wrong-sign (WS) $\Dz \to K^+\pi^-$ two-body decays\footnote{The inclusion of charge-conjugate processes is implicit unless stated otherwise.}.
The RS decay is dominated by a Cabibbo-favored (CF) amplitude, while the WS decay can proceed either through a doubly Cabibbo-suppressed (DCS) process, or 
through mixing ($\Dz \leftrightarrow \Dzb$), followed by a RS decay. The neutral $D$ flavor at production is tagged using the charge of the soft (low-momentum) pion $\pi^+_s$, in the decay $\Dstarp\to \Dz\pi^+_s$. In the limit of $x, y \ll 1$ and assuming \CP conservation, the decay-time-dependent ratio of WS-to-RS decay rates is approximated as~\cite{Artuso:2008vf,Bianco:2003vb,Burdman:2003rs,Blaylock:1995ay}:
\begin{equation}\label{eq:true-ratio}
R(t) \approx R_D+\sqrt{R_D}\ y'\ \frac{t}{\tau}+\frac{x'^2+y'^2}{4}\left(\frac{t}{\tau}\right)^2,
\end{equation}
where $t$ is the decay time, $\tau$ is the \Dz lifetime, $R_D$ is the ratio of DCS to CF decay rates, $x' \equiv x\cos\delta+y\sin\delta$, $y' \equiv y\cos\delta-x\sin\delta$,
      and $\delta$ is the strong phase difference between the DCS and CF amplitudes.

Allowing for \CP violation, the WS-to-RS yield ratios in Eq.~\eqref{eq:true-ratio} are written separately for \Dz and \Dzb as $R^+(t)$ and $R^-(t)$, respectively:
\begin{equation}\label{eq:true-ratio-pm}
R^\pm(t) \approx R^\pm_D+\sqrt{R^\pm_D}\ y'^{^\pm}\ \frac{t}{\tau}+\frac{x'^{2\pm}+y'^{2\pm}}{4}\left(\frac{t}{\tau}\right)^2.
\end{equation}
\CP violation in the WS decay amplitude (direct \CP violation) is characterized by the asymmetry parameter
$A_D \equiv (R^+_D-R^-_D)/(R^+_D+R^-_D)$. $A_D = 0$ if direct \CP symmetry is conserved.
Indirect \CP violation, which includes \CP violation either in mixing or in the interference between mixing and the decay amplitude,
 is characterized by the parameters $|q/p|$ and $\phi \equiv \arg(q/p)$. The mixing parameters are related by:
\begin{eqnarray}\label{eq:xyqop}
x'^{\pm} &=& \left(|q/p|\right)^{\pm 1} (x'\cos\phi\pm y'\sin\phi),  \nonumber \\
y'^{\pm} &=& \left(|q/p|\right)^{\pm 1} (y'\cos\phi\mp x'\sin\phi). 
\end{eqnarray} 
In the absence of indirect \CP violation, $|q/p|=1$, $\phi = 0$, and there will be no
difference between $(x'^{2+},\, y'^+)$ and $(x'^{2-},\, y'^-)$.

\section{Previous measurements}\label{sec:prevmeas}
First evidence for \Dz--\Dzb oscillation was reported in 2007 by the BaBar~\cite{Aubert:2007wf}, Belle~\cite{Staric:2007dt}, and CDF~\cite{Aaltonen:2007ac}
experiments. By 2009 the hypothesis of no oscillation was excluded with significance in excess of ten standard deviations by combining results from different experiments~\cite{HFAG}. In 2012 the \lhcb experiment reported a measurement of mixing parameters from the precursor to the present study and obtained the first observation from a single measurement with greater than five standard deviation significance~\cite{LHCb-PAPER-2012-038}, which has been recently confirmed by the CDF experiment~\cite{Aaltonen:2013pja}.

\section{\lhcb measurements}
The data used in this analysis comprise 1.0~\invfb of $\sqrt{s}=7$ TeV $pp$ collisions recorded during 2011, and 2.0~\invfb of $\sqrt{s}=8$ TeV $pp$ collisions recorded during 2012.
The \lhcb detector is a single-arm forward spectrometer covering the \mbox{pseudorapidity} range $2<\eta <5$~\cite{Alves:2008zz}. 

\begin{figure}[ht]
\centering
\includegraphics[width=0.4\textwidth]{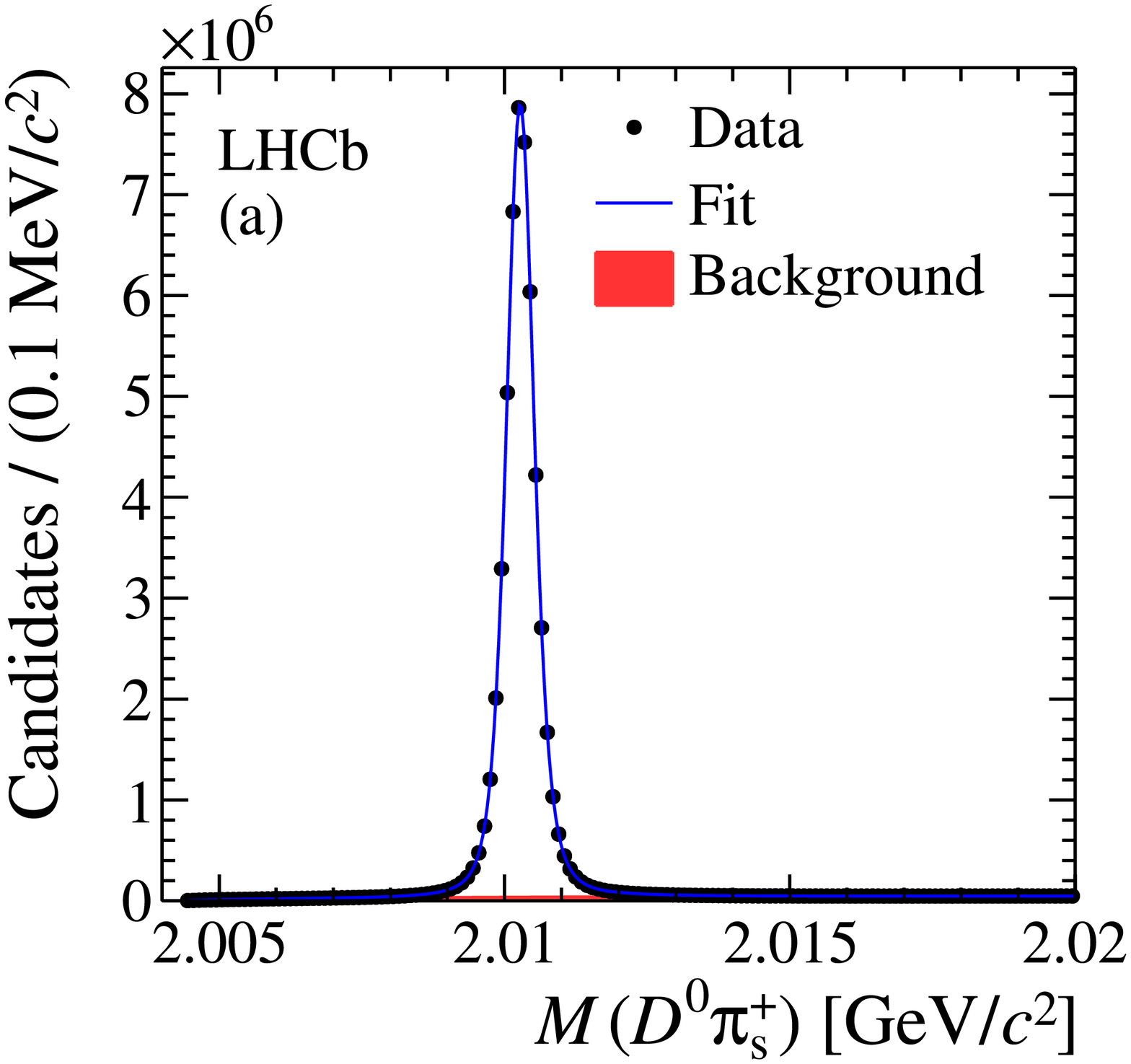}
\includegraphics[width=0.4\textwidth]{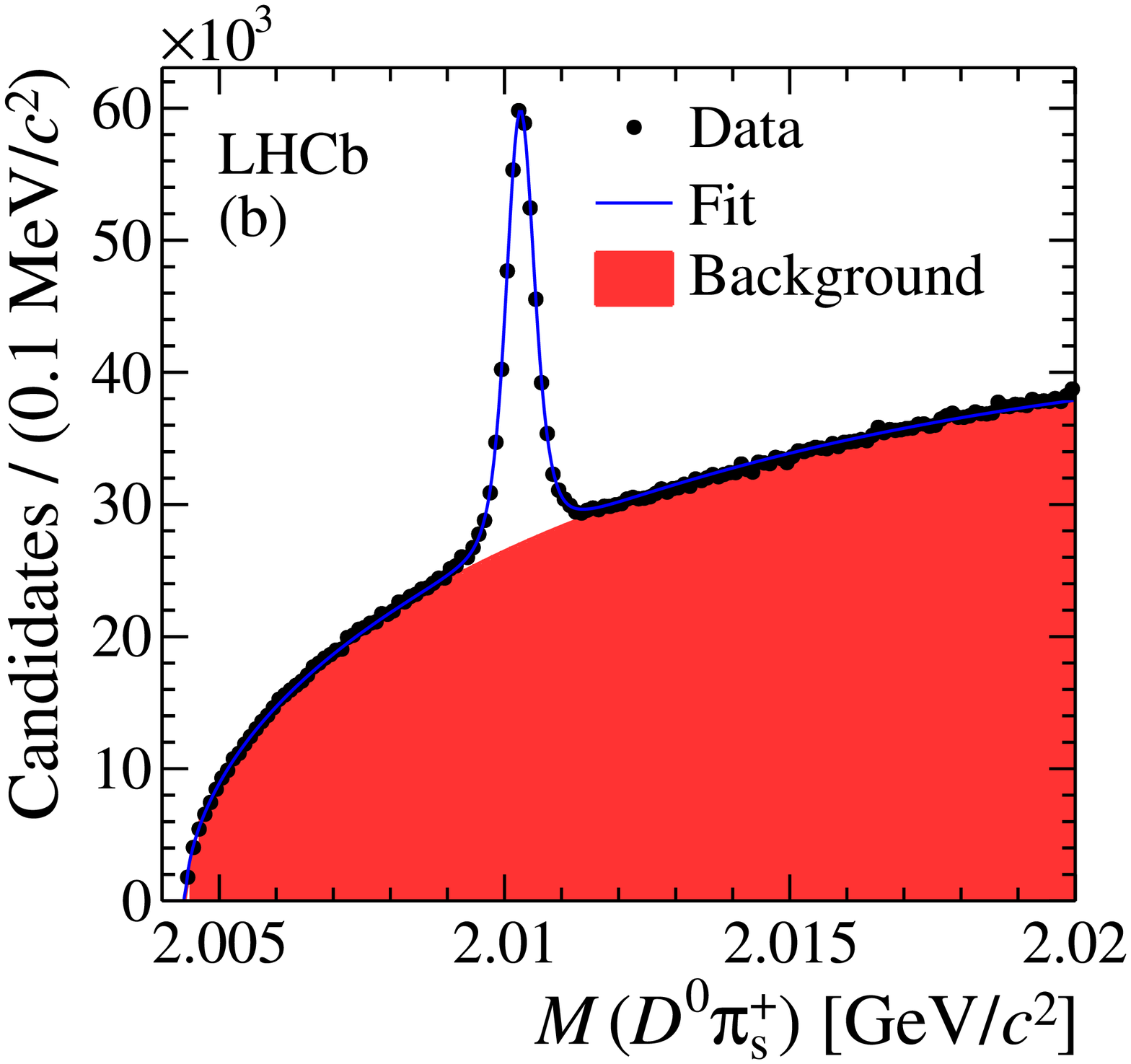}
\caption{\small Distribution of \MM  for selected (a) right-sign $\Dz\to K^-\pi^+$ and (b) wrong-sign $\Dz\to K^+\pi^-$ candidates.\label{fig:mass}}
\end{figure}

We select prompt $\Dstarp \to \Dz \pi^+_s$ decays that are consistent with production at 
the $pp$ collision point (primary vertex). The detailed event selection criteria are documented
in Ref.~\cite{thislhcbpaper}. The invariant mass of $K$ and $\pi$ from a \Dz candidate
is required to be within 24\,\massmev of the known \Dz mass, and the reconstructed 
$\Dz\pi^+_s$ mass, \MM, 
    is required to be lower than 2.02~\massgev.
The RS and WS signal yields are extracted by fitting the \MM 
distributions. The time-integrated \MM distributions of the selected RS and WS candidates and the associated fits are shown in Fig.~\ref{fig:mass}, where the depicted smooth background is dominated by favored $\Dzb\to K^+\pi^-$ decays associated with random $\pis^+$ candidates. In the fits, for a given \Dstar meson flavor, the signal shapes are common
to RS and WS decays, while the background shapes may differ.
In total, about 54 million signal RS decays and 0.23 million signal
WS decays are selected.

\begin{figure}[t!]
\centering
\includegraphics[width=0.5\textwidth]{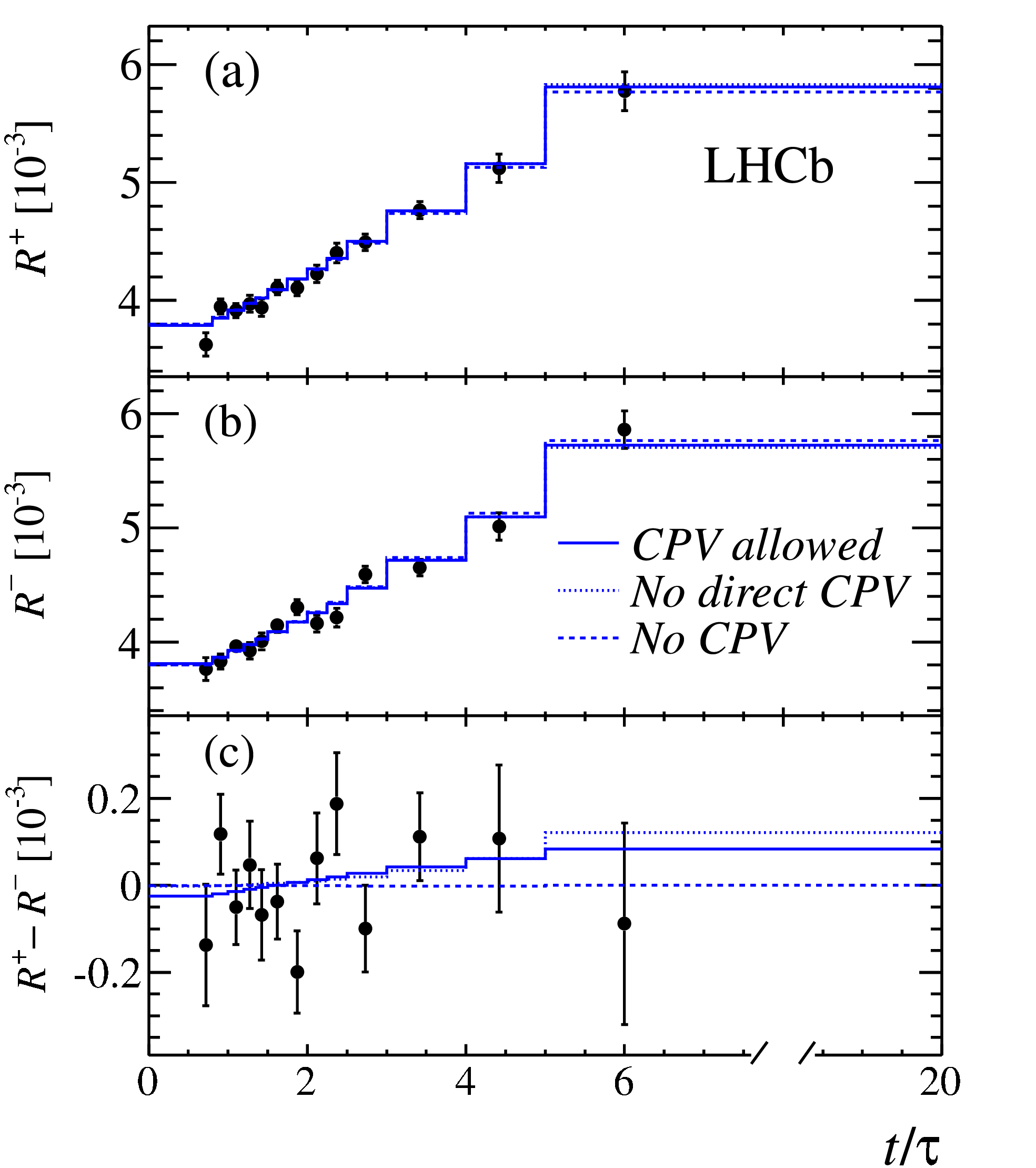}
\caption{\small Efficiency-corrected ratios of WS-to-RS yields for (a) \Dstarp decays, (b) \Dstarm decays, and (c) their differences as functions of decay time in units of \Dz lifetime. Projections of fits allowing for (dashed line) no \CP violation, (dotted line) no direct \CP violation, and (solid line) full \CP violation are overlaid. The abscissa of the data points corresponds to the average decay time over the bin; the error bars indicate the statistical uncertainties.\label{fig:finalResults}}
\end{figure}

The RS and WS samples for \Dz and \Dzb~mesons are each divided into 13 bins of 
\Dz decay time to compute decay-time-dependent WS-to-RS yield ratios. 
The ratios $R^+$ and $R^-$ observed in the \Dz and \Dzb samples and their differences are shown in Fig.~\ref{fig:finalResults}. 
These are corrected for the relative efficiencies $\epsilon_r^{\pm}$ to account for charge
asymmetries in reconstructing $\Kmp\pipm$ final states. 
The relative efficiencies are measured from data using the efficiency ratio
\begin{equation}\label{eq:effratio}
\epsilon_r^+\equiv 1/\epsilon_r^- \equiv \frac{\epsilon(\Kp\pim)}{\epsilon(\Km\pip)}=
\frac{N(\Dm\to \Kp\pim\pim)}{N(\Dp\to \Km\pip\pip)}\frac{N(\Dp\to \KS\pip)}{N(\Dm\to \KS\pim)}.
\end{equation}
With the asymmetry between \Dp and \Dm production rates canceled in the ratio, 
the $\Dpm\to \Kpm\pimp\pimp$ events are properly weighted to match the kinematics of the
$\Dpm\to\KS\pipm$ events. Similarly, these samples are weighted as functions of $K\pi$
momentum to match the RS momentum spectra. The charge asymmetry 
$A_{K\pi}\equiv (\epsilon_r^+-1)/(\epsilon_r^++1)$
is found to be in the range $0.8$--$1.2\%$ with $0.2\%$ precision, and independent of decay time. 

\begin{figure}[t!]
\centering
\includegraphics[width=0.48\textwidth]{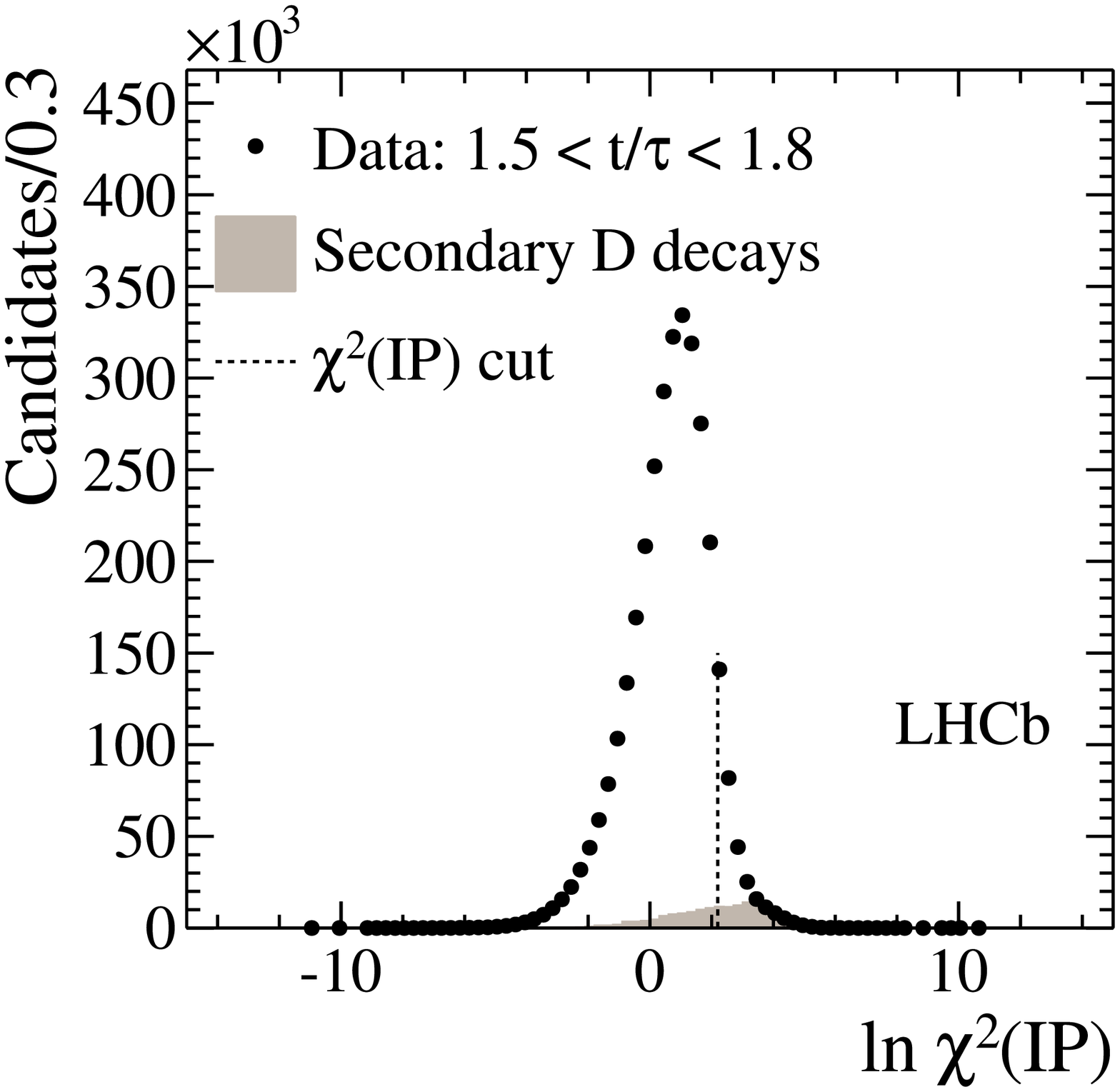}
\includegraphics[width=0.48\textwidth]{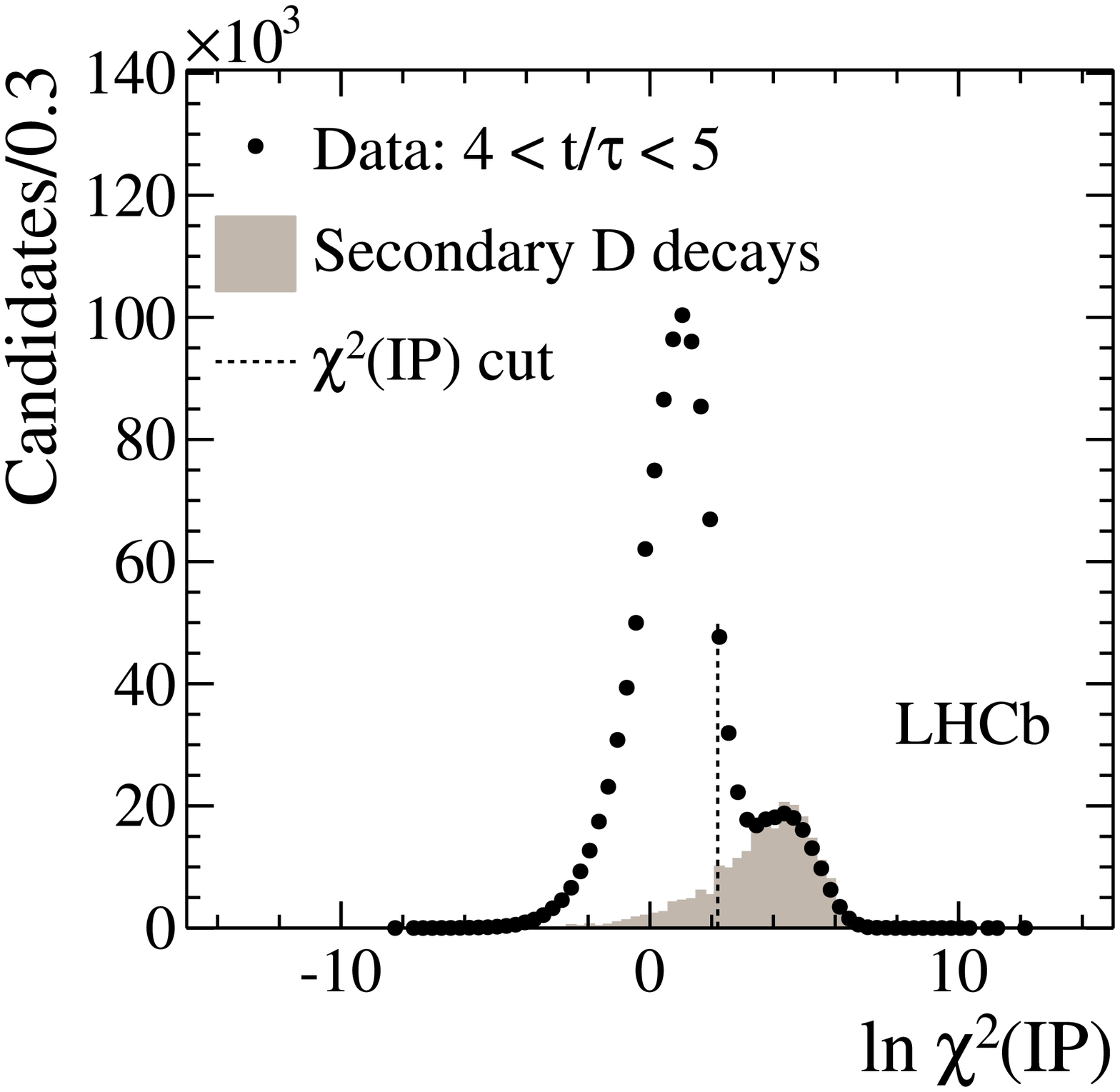}
\caption{Background-subtracted distributions of $\chi^2_{\mathrm{IP}}$ for RS decays in 
    two decay-time bins. The dashed line indicates the analysis selection requirement; the
hatched histogram represents the estimated secondary component.\label{fig:secfits}}
\end{figure}

\begin{figure}[t!]
\centering
\includegraphics[width=0.48\textwidth]{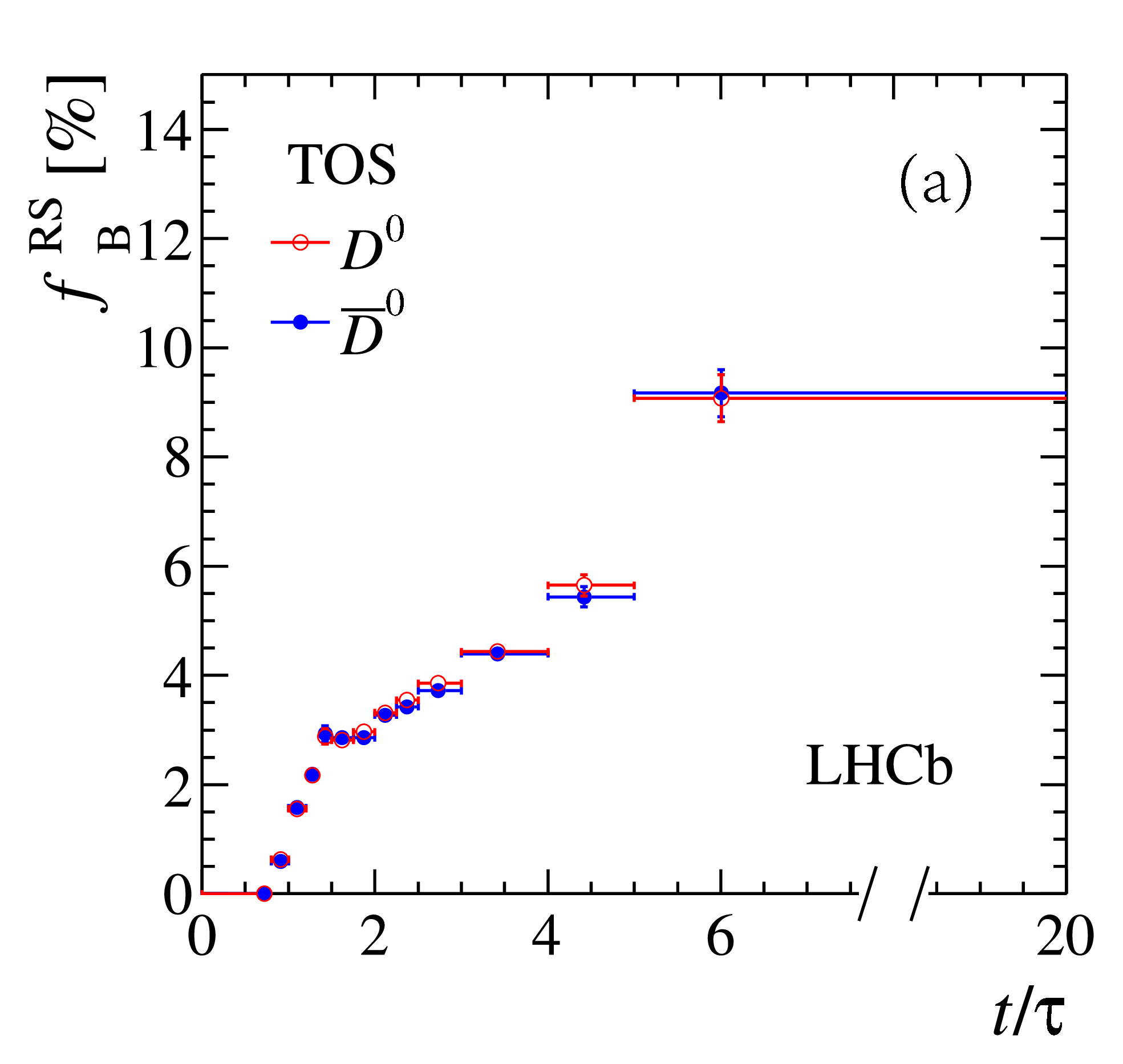}
\includegraphics[width=0.48\textwidth]{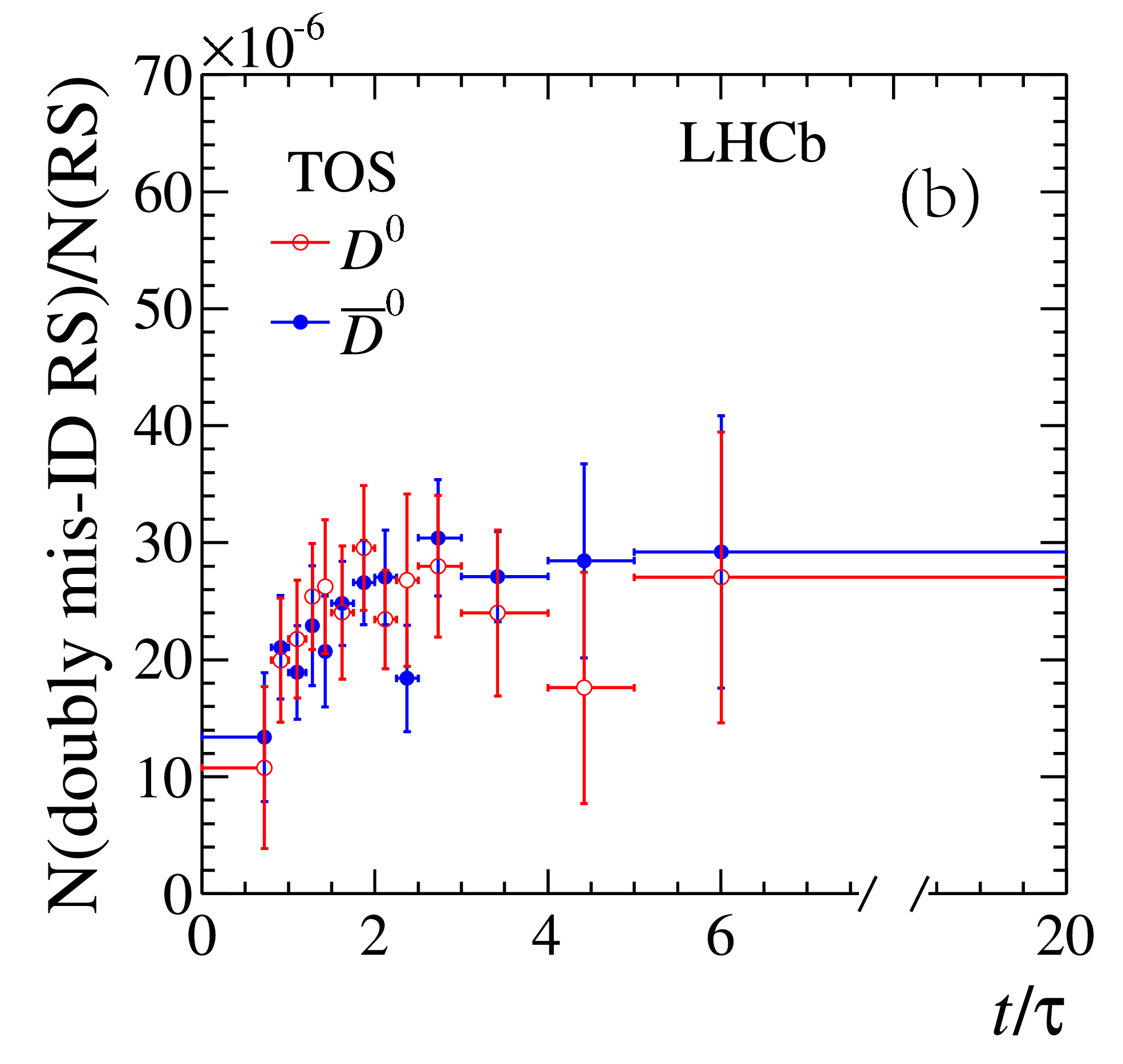}
\caption{Decay-time evolution of the contamination from (a) secondary $D$ decays and 
    (b) doubly misidentified RS candidates normalized to the RS signal yield,        
    for the data that meet the hardware trigger requirement (TOS),
        separately for \Dz and \Dzb decays.\label{fig:peakbkgs}}
\end{figure}

Charm mesons produced in $b$-hardron decays (secondary $D$ decays) are assigned with wrong decay time, and could
bias the measured WS-to-RS yield ratio. 
When the secondary component is not subtracted, the measured WS-to-RS yield ratio 
is written as $R(t)[1-\Delta_B(t)]$, where $R(t)$ is the ratio of the promptly 
produced candidates according to Eq.~\eqref{eq:true-ratio}, and $\Delta_B(t)$
is a time-dependent bias due to the secondary contamination.
Since $R(t)$ is measured to be monotonically non-decreasing~\cite{HFAG},
and the decay time for secondary decays is overestimated during reconstruction, 
$\Delta_B(t)$ can be bounded for all decay times as $0\leq \Delta_B(t) \leq f_B^{\mathrm{RS}}(t)\left[1-R_D/R(t)\right]$,
where $f_B^{\mathrm{RS}}(t)$ is the fraction of secondary decays in the RS sample at decay time $t$~\cite{hcp_proc_2012}. 
In this analysis, most of the secondary $D$ decays are removed
by requiring the $\chi^2$ of \Dz impact parameter with respect to the primary vertex, $\chi^2(\mathrm{IP})$,
   to be smaller than 9.
To determine the residual secondary decays, $f_B^{\mathrm{RS}}(t)$ is measured
by fitting the $\chi^2(\mathrm{IP})$ distribution of the RS \Dz candidates
in bins of decay time (see Fig.~\ref{fig:secfits}). The $\chi^2(\mathrm{IP})$
shape of the secondary component, and its dependence on decay time,
is also determined from data by studying the sub-sample of candidates
that are reconstructed, in combination with other tracks in the events,
as $B\to D^*\mu X$. Figure~\ref{fig:peakbkgs}~(a) shows
    the measured values of $f_B^{\mathrm{RS}}(t)$. 
We find that the secondary contamination is about 3\% fraction of signals, and has negligible
asymmetry when evaluated independently for \Dz and \Dzb decays.

Peaking background in \MM, that is not accounted
for in our mass fits, arises from \Dstar decays for which the $\pi_s^+$
is correctly reconstructed, but the \Dz~decay products are partially reconstructed or
misidentified. This background is suppressed
by the use of tight particle identification and $K\pi$
mass requirements. 
The dominant source of peaking
background leaking into our signal region is from RS $K\pi$
events which are doubly misidentified as a WS candidate.
This contamination is expected to have the same
decay time dependence of RS decays and, if neglected,
 would marginally affect the determination of the mixing
parameters, but lead to a small increase in the measured
value of $R_D$. From the events in the \Dz mass sidebands,
 we derive a bound on the possible time dependence of this
 background (see Fig.~\ref{fig:peakbkgs}~(b)). Contamination
 from peaking background due to partially reconstructed
 \Dz decays is found to be about 0.5\% of the WS signal candidates, and
 has negligible asymmetry when evaluated independently for \Dz and \Dzb decays.

\begin{table}
\centering
\caption{\small Results of fits to the data for different hypotheses on the \CP symmetry. The reported uncertainties include systematic effects.\label{tab:finalResults}}
\begin{footnotesize}
\begin{tabular}{l@{ [}lr@{\,$\pm$\,}l|l@{ [}lr@{\,$\pm$\,}l|l@{ [}lr@{\,$\pm$\,}l}
\hline\hline
\multicolumn{4}{c|}{Direct and indirect $CP$ violation} & \multicolumn{4}{c|}{no direct $CP$ violation} & \multicolumn{4}{c}{no $CP$ violation}\\\hline
 $ R_D $ & $10^{-3}$] & $3.568$ & $0.066$& $ R_D $ &$10^{-3}$] & $3.568$ & $0.066$ &$ R_D $ &$10^{-3}$] & $3.568$ & $0.066$ \\
 $ A_D $ &$10^{-2}$] & $-0.7$ & $1.9$ & $y'^+$&$10^{-3}$]     &  $ \phantom{-} 4.8 $ & $ 1.1 $ &$y'$  &$10^{-3}$]      &  $ \phantom{-}  4.8 $ & $ 1.0 $ \\
       $y'^+$ &$10^{-3}$]     &  $  \phantom{-} 5.1 $ & $ 1.4 $ & $x'^{2+}$ &$10^{-5}$]     &  $ \phantom{-} 6.4 $ & $ 5.5 $ &$x'^{2}$&$10^{-5}$]   &  $ \phantom{-} 5.5 $ & $ 4.9 $ \\
        $x'^{2+}$& $10^{-5}$]     &  $ \phantom{-} 4.9 $ & $ 7.0 $& $y'^-   $ &$10^{-3}$]     &  $ \phantom{-} 4.8 $ & $ 1.1$ & \multicolumn{2}{l}{$\chi^2/\text{ndf}$}  & \multicolumn{2}{l}{$\;\;\;\;86.4/101$} \\
         $y'^-$ & $10^{-3}$]     &  $ \phantom{-} 4.5 $ & $ 1.4$  & $x'^{2-}$&$10^{-5}$]     &  $ \phantom{-} 4.6 $ & $ 5.5 $  \\
          $x'^{2-}$& $10^{-5}$]     &  $ \phantom{-} 6.0 $ & $ 6.8 $ &\multicolumn{2}{l}{$\chi^2/\text{ndf}$}  & \multicolumn{2}{l|}{$\;\;\;\;86.0/99$}\\
\multicolumn{2}{l}{$\chi^2/\text{ndf}$}  & \multicolumn{2}{l|}{$\;\;\;\;85.9/98$} & \multicolumn{4}{l|}{}  \\
\hline\hline
\end{tabular}
\end{footnotesize}
\end{table}

\begin{figure*}[t]
\centering
\includegraphics[width=\textwidth]{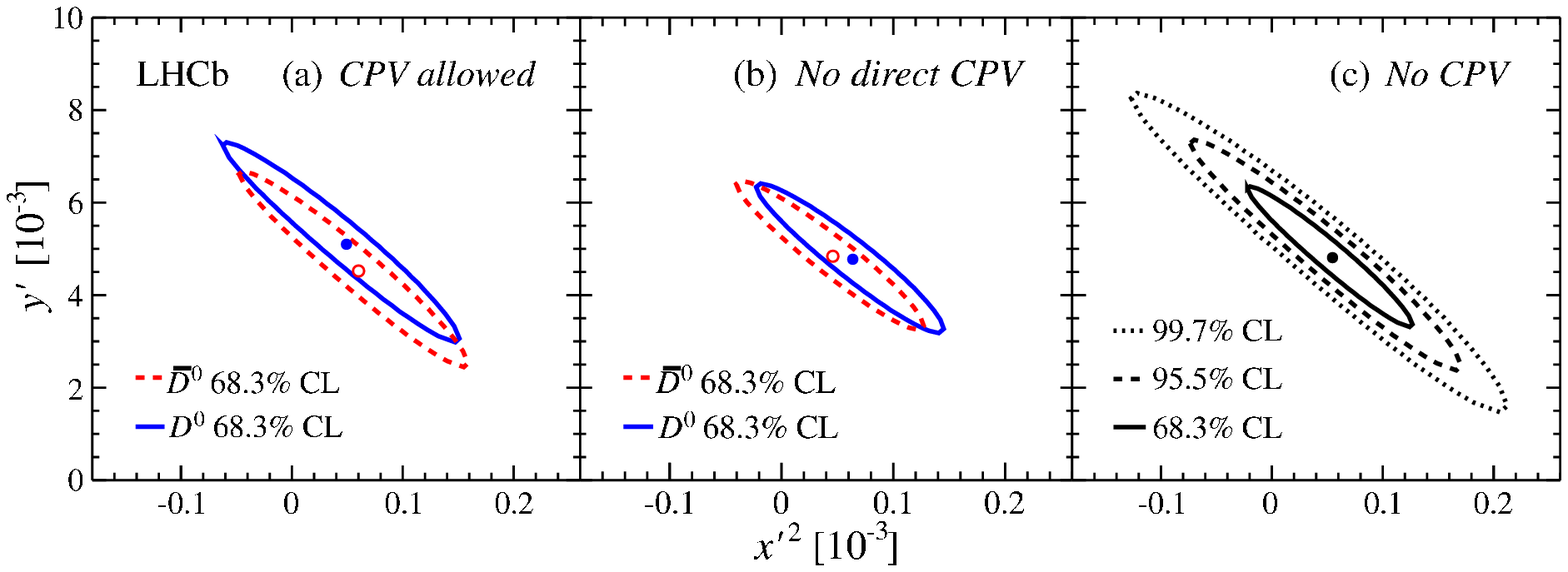}
\caption{\small Two-dimensional confidence regions in the $(x'^2,y')$ plane obtained (a) without any restriction on \CP violation, (b) assuming 
no direct \CP violation,  and (c) assuming \CP conservation. The dashed (solid) curves in (a) and (b) indicate the contours of the mixing parameters associated with \Dzb (\Dz) decays. The best-fit value for \Dzb (\Dz) decays is shown with an open (filled) point. The solid, dashed, and dotted curves in (c) indicate the contours of \CP-averaged mixing parameters at 68.3\%, 95.5\%, and 99.7\% confidence levels (CL), respectively. The best-fit value is shown with a point.
\label{fig:contours}}
\end{figure*}

Figure~\ref{fig:finalResults} shows that the WS-to-RS yield ratios from the data are fit three times. 
The first fit allows direct and indirect \CP violation, the second fit allows only indirect \CP violation by 
requiring a common value for $R_D$ in the \Dz and \Dzb samples, and the last
fit is a \CP-conserving fit that constrains all mixing parameters $(R_D,\, x'^{2},\, y')$ 
to be the same in both samples. 
The fit $\chi^2$ accounts for systematic effects due to the
decay-time evolution of the secondary $D$ decays and peaking background.
The fit results are shown in Table~\ref{tab:finalResults} and Fig.~\ref{fig:finalResults}, respectively. Figure~\ref{fig:contours} shows the central values and confidence regions in the $(x'^2,\, y')$ plane. 
The data are compatible with \CP symmetry.

\begin{figure*}[t!]
\centering
\includegraphics[width=0.48\textwidth]{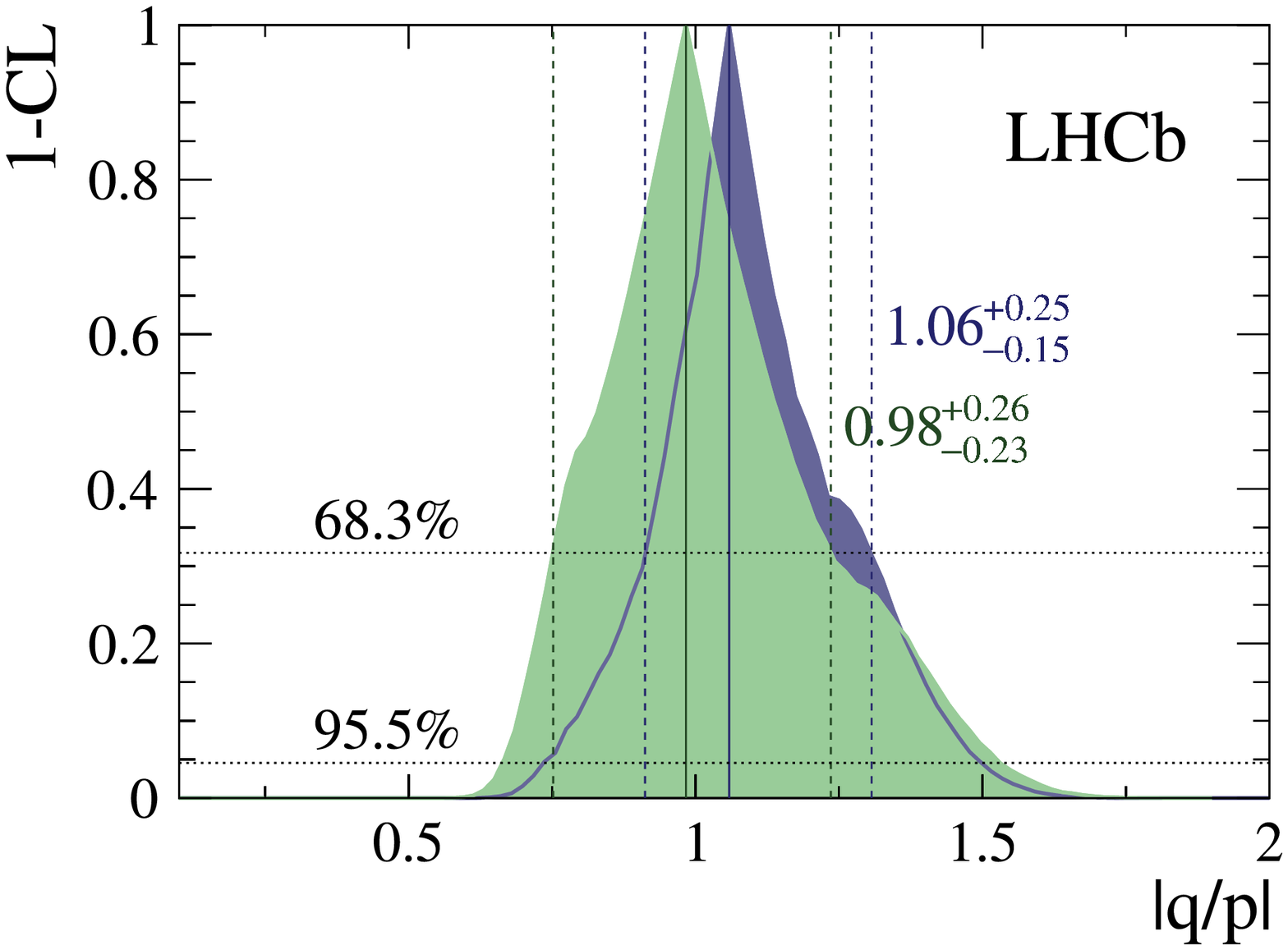}
\includegraphics[width=0.48\textwidth]{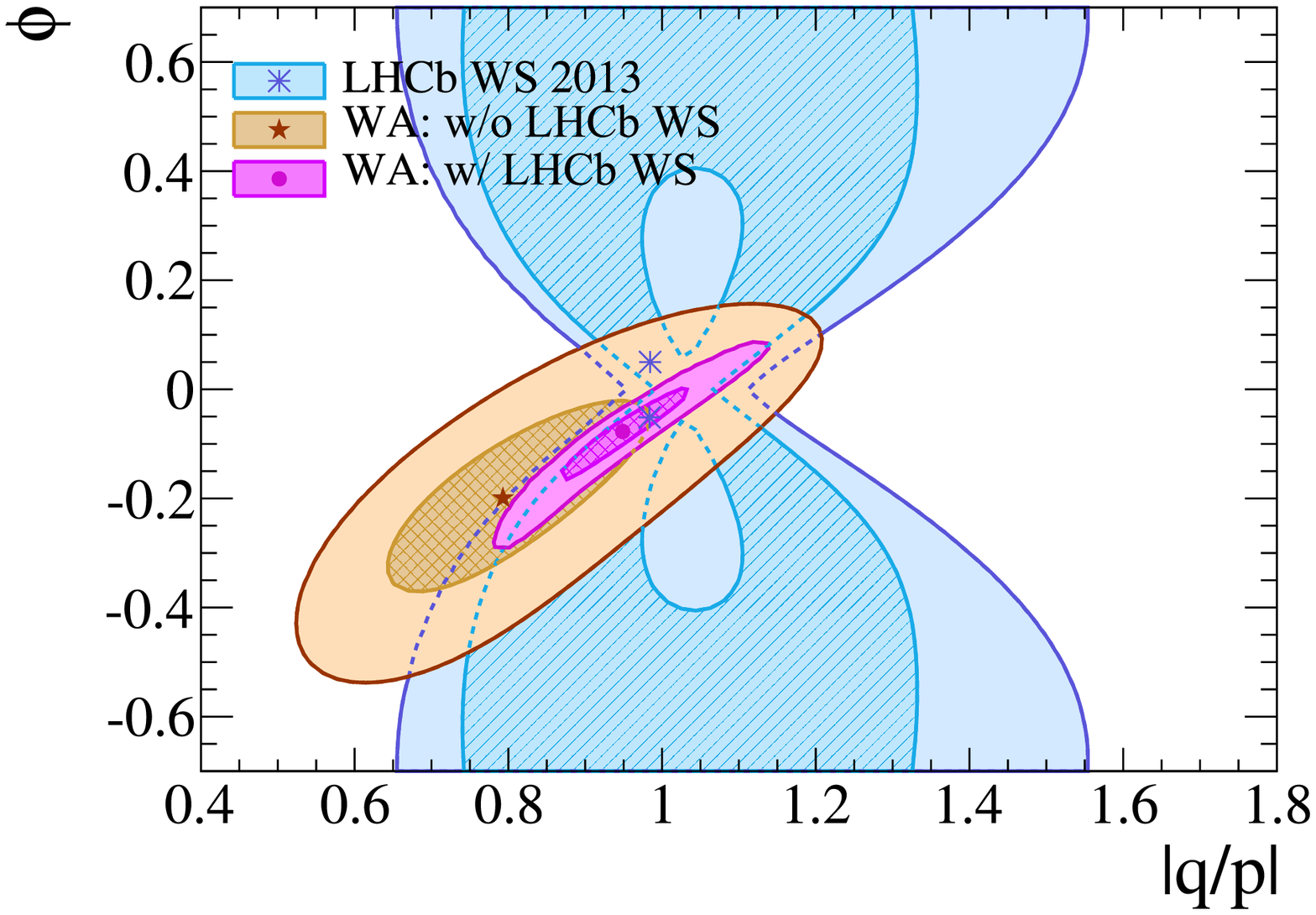}
\caption{Left: 
    (1 - CL) versus $|q/p|$ for the (green) direct and indirect \CP violation
and (blue) no direct \CP violation fit results. The reported numbers correspond to the best-fit
value and the uncertainties are computed using the respective 68.3\% CL intervals.
Right: 68.3\% and 95.5\% confidence regions
with (blue) only the \lhcb direct and indirect \CP violation allowed results,
 (brown) other measurements~\cite{HFAG} excluding the \lhcb WS results,
(magenta) other measurements including the \lhcb WS results.
These confidence regions are in 1D, so that the projection to the $|q/p|$ axis reproduces the 1D intervals.
 The \lhcb $A_{\Gamma}$ results~\cite{AGamma2013} are not taken into consideration.
\label{fig:qopcontours}}
\end{figure*}

From the fit results allowing for \CP violation, we build up a likelihood for $|q/p|$ using the relations of Eq.~\eqref{eq:xyqop}. 
Confidence intervals shown in Fig.~\ref{fig:qopcontours} are derived with a likelihood-ratio
ordering and assuming that the parameter correlations are independent of the true values of
the mixing parameters. At the 68.3\% CL, the magnitude of $q/p$ is determined to be $0.75 < |q/p|<1.24$
when any \CP violation is allowed, and $0.91<|q/p|<1.31$ for the case without direct \CP violation.
Figure~\ref{fig:qopcontours} demonstrates the power of the present results on constraining 
$|q/p|$ and $\phi$, 
when combined with other available measurements. In the limit that direct \CP violation is negligible,
and theoretical constraints such as the 
relationship $\phi = \tan^{-1}\left(\left(1-|q/p|^2\right)/\left(1+|q/p|^2\right)\right)$~\cite{Grossman:2009mn, Kagan:2009gb}
are applicable, the constraints on $|q/p|$
will be even more stringent~\cite{HFAG}.

The capability of the present results on constraining $|q/p|$ is also suggested 
by directly looking at the slopes observed in Fig.~\ref{fig:finalResults}.
Indirect \CP violation results in a time dependence of the efficiency-corrected
difference of WS-to-RS yield ratios.
In the limit of negligible direct \CP violation, and
$x'^{\pm}$, $y'^{\pm}$, and $\phi$ all very close to zero,
as suggested in Eq.~\eqref{eq:true-ratio-pm}
the slopes of the WS-to-RS yield ratios (Fig.~\ref{fig:finalResults}\,(a) and (b))
and the slope in the difference of yield ratios (Fig.~\ref{fig:finalResults}\,(c))
are proportional to $y'$ and $(|q/p|-|p/q|)y'$, respectively.
Within a span of about five decay-times, the slope in Fig.~\ref{fig:finalResults}\,(c)
is about 5\% of the individual slopes in Figs.~\ref{fig:finalResults}\,(a) and (b),
   and consistent with zero. 
Therefore, we expect $|q/p|$ to be constrained from one at a precision level
of a few percent at most.

\section{Summary}
Using $\Dstarp\to \Dz (\to K^+\pi^-) \pi^+$ decays reconstructed in 
3~\invfb of $pp$ collision data collected by the \lhcb experiment in 2011--2012,
    \Dz--\Dzb oscillation is studied with unprecedented level
    of precision. The observed mixing parameters $(R_D,\, x'^{2},\, y')$ assuming 
   \CP conservation are consistent with, $2.5$ times more precise than, and supersede the results based on a subset of the present data~\cite{LHCb-PAPER-2012-038}. Studying \Dz and \Dzb decays separately shows no evidence for \CP violation and provides the most stringent bounds on the parameters $A_D$ and $|q/p|$ from a single experiment. The present \lhcb \CP violation measurements also play an important role in constraining $|q/p|$ and $\phi$ when combined with 
   other measurements~\cite{HFAG}.



\Acknowledgements

 
\noindent We express our gratitude to our colleagues in the CERN
accelerator departments for the excellent performance of the LHC. We
thank the technical and administrative staff at the LHCb
institutes. We acknowledge support from CERN and from the national
agencies: CAPES, CNPq, FAPERJ and FINEP (Brazil); NSFC (China);
CNRS/IN2P3 and Region Auvergne (France); BMBF, DFG, HGF and MPG
(Germany); SFI (Ireland); INFN (Italy); FOM and NWO (The Netherlands);
SCSR (Poland); MEN/IFA (Romania); MinES, Rosatom, RFBR and NRC
``Kurchatov Institute'' (Russia); MinECo, XuntaGal and GENCAT (Spain);
SNSF and SER (Switzerland); NAS Ukraine (Ukraine); STFC (United
Kingdom); NSF (USA). We also acknowledge the support received from the
ERC under FP7. The Tier1 computing centres are supported by IN2P3
(France), KIT and BMBF (Germany), INFN (Italy), NWO and SURF (The
Netherlands), PIC (Spain), GridPP (United Kingdom). We are thankful
for the computing resources put at our disposal by
Yandex LLC (Russia), as well as to the communities behind the multiple open
source software packages that we depend on.


\begin{thebibliography}{99}

  
\bibitem{Artuso:2008vf} 
  M.~Artuso, B.~Meadows and A.~A.~Petrov,
  Ann.\ Rev.\ Nucl.\ Part.\ Sci.\  {\bf 58}, 249 (2008).
 
\bibitem{Bianco:2003vb} 
  S.~Bianco, F.~L.~Fabbri, D.~Benson and I.~Bigi,
  Riv.\ Nuovo Cim.\  {\bf 26N7}, 1 (2003).

\bibitem{Burdman:2003rs} 
  G.~Burdman and I.~Shipsey,
  Ann.\ Rev.\ Nucl.\ Part.\ Sci.\  {\bf 53}, 431 (2003).

  \bibitem{Blaylock:1995ay} 
  G.~Blaylock, A.~Seiden and Y.~Nir,
  Phys.\ Lett.\ B {\bf 355}, 555 (1995).
  
\bibitem{Petrov:2006nc} 
  A.~A.~Petrov,
  Int.\ J.\ Mod.\ Phys.\ A {\bf 21}, 5686 (2006).

\bibitem{Golowich:2007ka} 
  E.~Golowich, J.~Hewett, S.~Pakvasa and A.~A.~Petrov,
  Phys.\ Rev.\ D {\bf 76}, 095009 (2007).

\bibitem{Ciuchini:2007cw} 
  M.~Ciuchini, E.~Franco, D.~Guadagnoli, V.~Lubicz, M.~Pierini, V.~Porretti and L.~Silvestrini,
  Phys.\ Lett.\ B {\bf 655}, 162 (2007).

\bibitem{Aubert:2007wf} 
  B.~Aubert {\it et al.}  [BaBar Collaboration],
  Phys.\ Rev.\ Lett.\  {\bf 98}, 211802 (2007).

\bibitem{Staric:2007dt} 
  M.~Staric {\it et al.}  [Belle Collaboration],
  Phys.\ Rev.\ Lett.\  {\bf 98}, 211803 (2007).

\bibitem{Aaltonen:2007ac} 
  T.~Aaltonen {\it et al.}  [CDF Collaboration],
  Phys.\ Rev.\ Lett.\  {\bf 100}, 121802 (2008).

\bibitem{HFAG}
Heavy Flavor Averaging Group, Y.~Amhis {\it et~al.},
  arXiv:1207.1158 [hep-ex], updated
  results and plots available at
{\tt
  http://www.slac.stanford.edu/xorg/hfag/}.

\bibitem{LHCb-PAPER-2012-038} 
  R.~Aaij {\it et al.}  [LHCb Collaboration],
  Phys.\ Rev.\ Lett.\  {\bf 110}, 101802 (2013).

\bibitem{Aaltonen:2013pja} 
  T.~Aaltonen {\it et al.}  [CDF Collaboration],
  arXiv:1309.4078 [hep-ex].
  
\bibitem{Alves:2008zz} 
  A.~A.~Alves, Jr. {\it et al.}  [LHCb Collaboration],
  JINST {\bf 3}, S08005 (2008).


\bibitem{thislhcbpaper} 
  R.~Aaij {\it et al.}  [LHCb Collaboration],
  arXiv:1309.6534 [hep-ex].

\bibitem{hcp_proc_2012} 
  A.~Di Canto,
  EPJ Web Conf.\  {\bf 49}, 13009 (2013).

\bibitem{AGamma2013} 
  R.~Aaij {\it et al.}  [LHCb Collaboration],
 arXiv:1310.7201 [hep-ex].

\bibitem{Grossman:2009mn} 
  Y.~Grossman, Y.~Nir and G.~Perez,
      Phys.\ Rev.\ Lett.\  {\bf 103}, 071602 (2009).

\bibitem{Kagan:2009gb} 
  A.~L.~Kagan and M.~D.~Sokoloff,
      Phys.\ Rev.\ D {\bf 80}, 076008 (2009).

\end{thebibliography}
\end{document}